\documentclass{elsart3p}
\usepackage{graphicx}
\usepackage{amssymb}

\def\XXint#1#2#3{{\setbox0=\hbox{$#1{#2#3}{\int}$}
     \vcenter{\hbox{$#2#3$}}\kern-.5\wd0}}

\begin{document}
\begin{frontmatter}



\title{Self-similar asymptotics for a class of Hele-Shaw flows driven solely
by surface tension}


\author[label1]{Baruch Meerson},
\author[label2] {Pavel V. Sasorov},
\author[label1] {Arkady Vilenkin}
\address[label1]{Racah Institute  of  Physics, Hebrew University
of Jerusalem, Jerusalem 91904, Israel}
\address[label2]{Institute of Theoretical and Experimental
Physics, Moscow 117218, Russia}

\begin{abstract}
We investigate the dynamics of relaxation, by surface tension, of a family of
curved interfaces between an inviscid and viscous fluids in a Hele-Shaw cell. At
$t=0$ the interface is assumed to be of  the form  $|y|=A \, x^m$, where $A>0$,
$m \geq 0$, and $x>0$. The case of $0<m<1$ corresponds to a smooth shape, $m>1$
corresponds to a cusp, whereas $m=1$ corresponds to a wedge.  The inviscid fluid
tip retreats in the process of relaxation, forming a lobe which size grows with
time. Combining analytical and numerical methods we find that, for any $m$, the
relaxation dynamics exhibit self-similar behavior. For $m\neq 1$ this behavior
arises as an intermediate asymptotics: at late times for $0\leq m<1$, and at
early times for $m>1$. In both cases the retreat distance and the lobe size
exhibit power law behaviors in time with different dynamic exponents, uniquely
determined by the value of $m$. In the special case of $m=1$ (the wedge) the
similarity is exact and holds for the whole interface at all times $t>0$, while
the two dynamic exponents merge to become $1/3$. Surprisingly, when $m\neq 1$,
the interface shape, rescaled to the local maximum elevation of the interface,
turns out to be universal (that is, independent of $m$) in the similarity
region. Even more remarkably, the same rescaled interface shape emerges in the
case of $m=1$ in the limit of zero wedge angle.
\end{abstract}

\begin{keyword}
unforced Hele-Shaw flow \sep surface tension \sep relaxation \sep scaling \sep
power law \sep self-similarity
\end{keyword}

\end{frontmatter}


\section{Introduction}
\label{intro} Consider a curved interface between a low-viscosity fluid (for
example, water) and a high-viscosity fluid (for example, oil) in a large
horizontal Hele-Shaw cell. If the system is unforced, the interface undergoes
relaxation by surface tension, ultimately approaching either a straight line, or
a circle (or breaking into several domains which then become circles). This free
boundary problem is hard to solve, as the governing equations (see below) are
non-local. This is especially true when the interface initially has a complex
shape, as observed in numerous experiments when the viscous fluid is displaced
by the inviscid fluid in radial geometry, see Ref. \cite{Sharon} and references
therein. The initial shape complexity is caused here by the viscous fingering
instability that develops during the preceding \textit{forced} stage of the flow
\cite{ST,Paterson}. The (strongly) forced Hele-Shaw flow is a standard paradigm
in fluid dynamics and nonlinear dynamics
\cite{Langer1,Kadanoff,Kessler,Casademunt}. The (small) surface tension there is
usually invoked in order to regularize the otherwise singular dynamics on small
scales. We are interested in this paper in an \textit{unforced} flow, where
surface tension is the \textit{only} driving mechanism. Here is the formulation
of the unforced Hele-Shaw (UHS) flow model that we will be dealing with
throughout this paper. Let one fluid have negligible viscosity, so that the
pressure in it is uniform. The velocity of the viscous fluid is
$\mathbf{v}\,(\mathbf{r},t)=-(H^2/12\mu) \,\nabla p\,(\mathbf{r},t)$, where $p$
is the pressure, $\mu$ is the dynamic viscosity, and $H$ is the plate spacing
\cite{ST,Paterson,Langer1,Kadanoff}. Therefore, the interface speed is
\begin{equation}\label{speed}
  v_n = - (H^2/12 \mu) \,\partial_n p\,,
\end{equation}
where index $n$ denotes the components of the vectors normal to the interface
and directed from the inviscid fluid to the viscous fluid, and $\partial_n p$ is
evaluated at the corresponding point of the interface $\gamma$. The viscous
fluid is incompressible. Therefore, its pressure is a harmonic function:
\begin{equation}\label{Laplace}
  \nabla^2 p =0\,.
\end{equation}
The Gibbs-Thomson relation yields a boundary condition at the interface:
\begin{equation}\label{jump2}
  p\,|_{\gamma} = (\pi/4)\,\sigma {\mathcal{K} }\,,
\end{equation}
where $\sigma$ is surface tension, and ${\mathcal{K}}$ is the local curvature of the
interface, positive when the inviscid region is convex outwards. Finally, as the
flow is unforced, we demand
\begin{equation}\label{external}
\nabla p\, = 0 \;\;\;\mbox{at}\;\;\;\mathbf{r}\to \infty\,.
\end{equation}
Equations (\ref{speed})-(\ref{external}) define the UHS problem (see Refs.
\cite{Constantin1,Almgren,CLM} and references therein for a more detailed
discussion). The UHS model gives an instructive example of non-local
area-preserving curve-shortening dynamics.

The UHS flow (\ref{speed})-(\ref{external}) is not integrable.
Moreover, until
recently even no \textit{particular} analytic solutions to this class of flows
had been found, except for the simple solutions provided by a linear stability
analysis of a single, slightly deformed flat or circular interface
\cite{linear}. Recently, some analytic solutions have been obtained for two
special initial interface shapes. In the first of them, the inviscid fluid
domain at $t=0$ has the form of a half-infinite stripe \cite{VMS}. As time
progresses, the tip of the stripe retreats and develops a lobe. At long times,
the growing lobe approaches a self-similar shape, whereas the lobe size and
retreat distance follow a power law in time with different dynamic exponents:
$1/5$ and $3/5$, respectively.

In the second case the assumed form of the inviscid fluid was a wedge
\cite{GMV}. As this initial condition, and Eqs. (\ref{speed})-(\ref{external}),
do not introduce any length scale into the problem, the solution is self-similar
at \textit{all} times, with a single dynamic exponent $1/3$ \cite{GMV}. The
scale-invariant interface shape in this case is given by the solution of an
unusual inverse problem of potential theory. Gat \textit{et al.} \cite{GMV}
solved this problem perturbatively for an almost flat wedge, and numerically for
several values of the wedge angle.

The results of Refs. \cite{VMS,GMV} suggest that the values of dynamic
exponents, and other attributes of the self-similar asymptotics, are determined
by the initial shape of the retreating edge of the inviscid fluid domain, while
the two solutions obtained in Refs. \cite{VMS,GMV} are particular members of a
broader family of solutions. The results of the present paper confirm this
scenario. We consider here a more general, power-law shape $|y|=A \, x^m$, where
$A>0$, $m \geq 0$, and $x>0$, and show that, for \textit{any} $m \geq 0$, the
relaxation dynamics exhibit self-similar intermediate asymptotics: a late-time
asymptotics for $0 \leq m<1$ and an early-time asymptotics for $m>1$. The
retreat distance and the lobe size show, for any $m$, a power law behavior in
time with exponents and pre-factors uniquely determined by $m$. The case of
$m=1$, investigated in Ref. \cite{GMV}, is special: here the self-similarity is
exact and occurs for \textit{all} times $t>0$, while the two dynamic exponents
merge and become equal to $1/3$. Surprisingly, at $m\neq 1$ the interface shape,
rescaled to the local maximum elevation of the interface, turns out to be
universal (that is, independent of $m$) in the similarity region. Even more
remarkably, the same rescaled interface shape emerges in the case of $m=1$ in
the limit of zero wedge angle.

Here is a layout of the rest of the paper.  In Section II we generalize to an
arbitrary $m\ge 0$ the approach, suggested by Vilenkin \textit{et al.}
\cite{VMS} for a half-infinite stripe ($m=0$). We present there a simple
asymptotic scaling analysis that predicts (i) the dynamic exponents of the
self-similar part of the flow, (ii) the exponent of the power-law tail of the
scale-invariant shape function of the interface, and (iii) the validity range of
the scaling behavior at $0\leq m <1$ and $m>1$. In Section III we report the
results of a numerical solution of (\ref{speed})-(\ref{external}) for the cases
of $m=1/4$, $1/2$ and $5/4$, compare them with our theoretical predictions and
report the universality of the rescaled interface shape. In addition, we present
in Section III our new numerical results for small-angle wedges. A brief
discussion and summary are presented in Section IV.

\section{Interface dynamics: theoretical predictions} \label{theory}

\begin{figure}[ptb]
\hspace{0.5cm}
\includegraphics[width=6.5 cm,clip=]{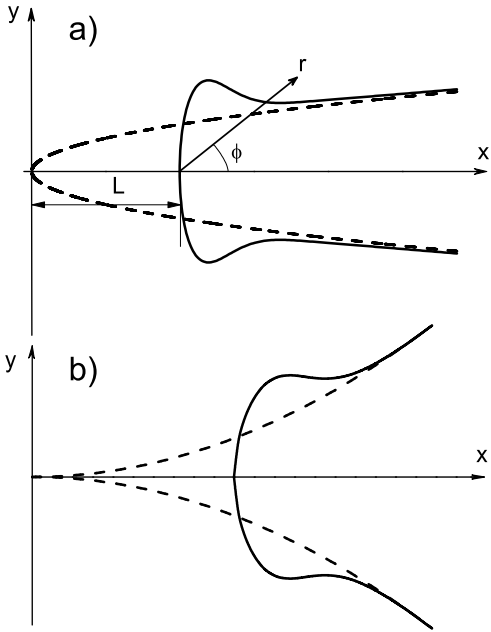}
\caption{A schematic setting for the interface dynamics for $0\le m <1$ (a) and
$m>1$ (b).} \label{picture}
\end{figure}

Let at $t=0$ the interface shape be $|y|=A \, x^m$, where $A>0$, $m\geq 0$, and
$x>0$. The case of $0< m<1$ corresponds to smooth shapes, $m>1$ corresponds to a
cusp,  see Fig.~\ref{picture}, while $m=1$ corresponds to a wedge. The parameter
$A$ has the dimension of \textit{length}$^{1-m}$. This implies that the case of
$m=1$, investigated by Gat \textit{et al.} \cite{GMV}, is special, as the
parameter $A$ does not introduce any length scale there. We assume in this
Section that $m \neq 1$ and measure the distance in units of $\Delta \equiv
A^{1/(1-m)}$, the time in units of $\tau=48 \mu \Delta^3/(\pi \sigma H^2)$, and
the pressure in units of $p_0=\pi \sigma/(4\Delta)$. In the rescaled variables
the interface shape is $|y|=x^m$, while Eqs. (\ref{speed}) and (\ref{jump2}) are
parameter-free:
\begin{equation}\label{speed1}
v_n = - \partial_n p
\end{equation}
and
\begin{equation}\label{jump3}
p\,|_{\gamma} = {\mathcal{K}}\,.
\end{equation}

Because of the finite surface tension, the pressure gradient in the viscous
fluid is largest at the tip, so the tip will retreat along the x-axis and
develop a lobe which size will grow with time. Let us assume (and then test
numerically) that the evolving interface can be fully characterized by
\textit{two} time-dependent length scales: the curvature radius, $R(t)$, and the
retreat distance, $L (t)$,  of the tip. In the leading order this assumption
implies a similarity ansatz, in the moving frame $x_1=x-L(t)$, for the interface
elevation $y=h(x,t)$ in the lobe region and region II (see below):
\begin{equation}\label{Ansatz}
h_s(x_1,t)=R(t)\, \Phi\left[\frac{x_1}{R(t)}\right]\,.
\end{equation}
In the scaling regime, the dynamic length scales  $L(t)$ and $R(t)$ exhibit a
power law behavior:
\begin{equation}\label{fullpowerlaws}
    L(t)=a\,
t^{\alpha} \;\;\;\mbox{and}\;\;\;  R(t)=b\, t^{\beta}\,,
\end{equation}
where the exponents $\alpha$ and $\beta$ and coefficients $a$ and $b$ are
$m$-dependent. The objectives of this work is to determine analytically
$\alpha$, $\beta$ the exponent of the power-law tail of the scale-invariant
shape function of the interface $\Phi\left[x_1/R(t)\right]$, and the validity
range of the scaling behavior at $0\leq m <1$ and $m>1$. These analytic findings
will be verified and complemented by our numerical solutions which, in addition,
provide the whole scale-invariant shape function $\Phi\left[x_1/R(t)\right]$ and
the coefficients $a$ and $b$ of the power laws of $R(t)$ and $L(t)$ for four
different values of $m$, in Sec. III.

A simple asymptotic scaling analysis for $m\neq 1$ is possible because of the
following property of the initial shape of the inviscid fluid region: it has a
flat region at either very large distances $x \gg 1$ (for $0\leq m <1$), or very
small distances $x\ll 1$ (for $m>1$). We will see that, as a result, the scaling
regime will hold at $t \gg 1$ for $0\leq m <1$ and at $t \ll 1$ for $m>1$. The
calculations are very similar to those of Ref. \cite{VMS}, where the particular
case of $m=0$ was considered. Introduce for a moment the system of polar
coordinates $r$ and $\phi$ with the origin at the moving tip of the interface,
as shown in Fig.~\ref{picture}a. In view of the Gibbs-Thomson condition
(\ref{jump3}), $p$ must vanish, in the leading order, at the flat region, that
is at $\phi \to 0$ and $\phi \to 2\pi$. This corresponds either to the region
$1\ll R(t) \ll r$ for $0\leq m<1$, or to the region $R(t) \ll r \ll 1$ for
$m>1$. Furthermore, in view of the same Eq. (\ref{jump3}), $p={\mathcal{K}} \sim
1/R(t)$ in the lobe region (for definiteness, at $\phi= \pm \pi/2$). Therefore,
the leading term of the far-field multipole expansion of $p$ \cite{Jackson} can
be written as
\begin{equation}
p (r, \phi, t) = C_m \left[R(t)\,r\right]^{-1/2}\, \sin(\phi/2) , \label{B1}
\end{equation}
where $C_m ={\mathcal{O}}(1)$. Having demanded the boundary condition
(\ref{jump3}) in Eq. (\ref{B1}), we stretched the validity region of Eq.
(\ref{B1}), $r\gg R(t)$, toward $r\sim R(t)$, but this can only affect the value
of constant $C_m$ that our theory cannot give anyway. Now we can employ Eqs.
(\ref{speed1}) and (\ref{B1}) and estimate the normal component of the interface
speed in the far field region. For the upper interface of the far field region
\begin{equation}
\label{v_n} v_n = -\frac{1}{r} \frac{\partial p}{\partial \phi}(\phi\to 0) =
-\frac{C_m}{2 R^{1/2}(t)\, r^{3/2}}\,.
\end{equation}
At this point we return to the Cartesian coordinates $x_1, y$. In the far field
region, that is at $1 \ll R(t) \ll x_1$ for $0\leq m<1$ or $R(t) \ll x_1 \ll 1$
for $m>1$, the quantity $\partial h(x_1,t)/\partial t$ is given by Eq.
(\ref{v_n}), and we obtain
\begin{equation}\label{integral}
h(x_1,t) -x^m = \int_0^t \frac{C_m\,dt^\prime}{2 R^{1/2}(t^\prime)\,
x_1^{3/2}}\,.
\end{equation}
As we will check \textit{a posteriori}, $L(t)$ is always much greater than
$R(t)$ in the scaling regime. How can we simplify the calculation of the
intergal in Eq.~(\ref{integral})? The far field region $x_1\gg R(t)$ can be
divided into two sub-regions: $x_1\gg L(t)$ (region I) and $R(t) \ll x_1 \ll
L(t)$ (region II, or neck). In region I we can, in the leading order, take
$x_1^{3/2}$ out of the integral and arrive at
\begin{equation}\label{veryfar}
   h(x_1,t) -x^m\simeq \frac{C_m}{2\,
   x_1^{3/2}}\int_0^t\frac{dt^\prime}{R^{1/2}(t^\prime)}\, \sim
   \frac{t}{x_1^{3/2}\,R^{1/2}(t)}\,,
\end{equation}
where we have assumed that $R(t)$ is a power of $t$, and disregarded the
coefficient $C_m={\mathcal{O}}(1)$.

In region II at fixed $x$, the main contribution to the integral in
Eq.~(\ref{integral}) comes from times close to $t$, so that $x_1(t)/\dot{L}(t)
\ll t-t^{\prime} \ll t\,$. Indeed, we can expand $x_1(t^\prime) =
x_1(t)+\dot{L}(t) (t-t^\prime)+ \dots\,$ and, in the leading order, neglect
higher order terms. The effective time interval for the integration is
$(t-\delta t^{\prime}, t)$, where $\delta t^{\prime} \sim x_1(t)/\dot{L}(t)$.
Furthermore, $R^{1/2}(t^\prime)$ can be evaluated at $t^{\prime}=t$, as its
variation  on the time interval $(t-\delta t^{\prime}, t)$ is negligible. Then,
extending the lower limit of the integral to $-\infty$ and calculating the
remaining elementary integral, we obtain
\begin{equation}\label{far}
h(x_1,t) -x^m\sim \frac{C_m}{R^{1/2}(t)\,\dot{L}(t)x_1^{1/2}}\,,
\end{equation}
where the factor $C_m={\mathcal{O}}(1)$  is in excess of accuracy and can be
disregarded. Now we can estimate the contributions of regions I and II to the
area gain $\delta A$ in the far field region. In region I (correspondingly, II)
the main contribution to the integral over $x_1$ comes from the lower
(correspondingly, upper) limit of integration. Therefore, we integrate Eq.
(\ref{veryfar}) over $x_1$ from, say, $2 L(t)$ to infinity, and Eq. (\ref{far})
from $R(t)$ to $2 L(t)$. The results are:
\begin{equation}\label{regionI}
 \delta A_I (t) \sim \frac{t}{L^{1/2}(t)\,R^{1/2}(t)}\;\;\;\;\;\;\mbox{in region I}\,,
\end{equation}
and
\begin{equation}\label{regionII}
\delta A_{II} (t) \sim
\frac{L^{1/2}(t)}{\dot{L}(t)\,R^{1/2}(t)}\;\;\;\;\;\;\mbox{in region II}\,.
\end{equation}
As, by assumption, $L(t)$ is a power law, $\delta A_I$ is comparable to $\delta
A_{II}$. As we will check shortly, the contribution to the area of the lobe
region itself, $\delta A_R\sim R^2(t)$, is negligible compared to $\delta A_I$
and $\delta A_{II}$ as long as we are in the scaling regime ($t \gg 1$ for $0\le
m<1$ or $t\ll 1$ for $m>1$).

Now we employ the exact integral of motion of the system: the area conservation
of each of the fluids. The area loss because of the retreat [which is equal to
$L(t)^{m+1}/(m+1) \sim L(t)^{m+1}$] must be equal, in the leading order, to the
area gain in the far field region. This follows
$$L(t)^{m+1} \sim \delta A_I(t) \sim
\delta A_{II}(t)\,,$$ which yields a relation between the two dynamic length
scales $R(t)$ and $L(t)$. Another relation between these two quantities follows
from Eq.~(\ref{B1}). We obtain $V_{l} \sim -\partial p/\partial r \left[r \sim
R(t), \phi \simeq \pi\right] \sim R^{-2}(t)$, and demand
$$\dot{L}(t)\sim R^{-2}(t)\,.$$
These two relations immediately yield the dynamic exponents $\alpha$ and
$\beta$:
\begin{equation}
\alpha=\frac{3}{4m+5} \;\;\;\;\;\;\mbox{and}\;\;\; \;\;\;\beta
=\frac{2m+1}{4m+5}\,. \label{scalings2}
\end{equation}
Once the scaling relations for $L(t)$ and $R(t)$ are found, we can calculate [up
to $m$-dependent numerical pre-factors ${\mathcal{O}}(1)$], additional
quantities. For example, the interface elevation in region I becomes
$$h(x,t)-x^m \sim t^{\frac{3(2m+3)}{2(4m+5)}} x^{-3/2}\,,$$
see Eq.~(\ref{veryfar}). In region II, see Eq.~(\ref{far}), we obtain
$$h(x,t)-x^m \sim t^{\frac{3 (2m+1)}{2 (4m+5)}} x^{-1/2}\,.$$
Importantly, region II belongs to the similarity region, as was first observed
in Ref. \cite{VMS} in the case of $m=0$. In this region $\Phi(\xi)\sim
\xi^{-1/2}$: a universal ($m$-independent) power law of the similarity variable
$\xi=x_1/R(t)$. The presence of the decreasing asymptote $\Phi(\xi)\sim
\xi^{-1/2}$ implies that, for any $m\neq1$, the shape function $\Phi(\xi)$ must
have a local maximum.

Once we obtained the solution, we can check it for self-consistency with all the
assumptions we made. First, it can be easily checked that, in the scaling regime
$t\gg 1$ (for $0\leq m<1$) or $t\ll 1$ (for $m>1$) we have $L(t)\gg R(t)$, as we
assumed. Now, the lobe area $\delta A_R \sim R^2(t)$ grows with time as  $\sim
t^{2(2m+1)/(4m+5)}$. This value is indeed much less than $\delta A_I(t)\sim
\delta A_{II}(t)\sim t^{3(m+1)/(4m+5)}$ in the scaling regime. Furthermore, the
time-dependent interface elevation in the lobe region $R(t) \sim
L(t)^{(2m+1)/3}$ is much larger, in the scaling regime, than the initial
interface elevation at the moving tip, which is $L(t)^m$. This strong inequality
serves as a necessary condition for the assumed similarity asymptotics
(\ref{Ansatz}) as a leading order description.

\section{Numerical solution} \label{numerics}

In order to test the predicted dynamic exponents $\alpha$ and $\beta$ and verify
the presence of the self-similar regime, we solved the problem numerically for
three different values of parameter $m$. In addition, we returned to the case of
$m=1$, previously considered in Ref. \cite{GMV}, and solved it numerically for
small values of the wedge angle.

\subsection{Numerical method}

Our numerical algorithm \cite{VM} employs a variant of the boundary integral
method for an exterior Dirichlet problem formulated for a singly connected
domain, and explicit tracking of the contour nodes. The harmonic potential is
represented as a potential produced by a dipole distribution with an \textit{a
priori} unknown density $\mathbf{D}$ on the contour. The dipole density
$\mathbf{D}$ is found numerically from an  integral equation which is a
modification of the well-known jump relation of the potential theory \cite{GGM}.
Computing another integral of this (already found) dipole density, one obtains
the harmonic conjugate function, whose derivative along the contour yields, by
virtue of the Cauchy theorem, the normal velocity of the interface.

We used a piecewise constant function for a discrete approximation of
$\mathbf{D}$  and a piecewise linear function for discretizing  the interface.
The integral entering the integral equation is represented as a sum of
$\mathbf{D}$ multiplied by a kernel which is integrated analytically between two
neighboring nodes. This approximation was previously suggested for the inner
problem \cite{Munck}. We found that it is also efficient in the outer problem in
the following cases: (i) for a long and slender domain, (ii) in the vicinity of
the cusp at $m>1$, and (iii) for a wedge ($m=1$) with a small angle. The
numerical approximation is described in detail in Ref. \cite{VM}. The method
requires an inhomogeneous grid with a small spacing in regions of high curvature
of the contour, and we used a grid with spacing exponentially growing with the
distance from the interface's tip. The number of grid nodes was reduced as the
interface's perimeter decreased, and the curvature radius of the tip increased.

The shape of the numerical interface at $t=0$ is determined by the following
parameters: the exponent $m$, the domain size $\Lambda>0$, and the cutoff
parameter $\varepsilon>0$ that was used for $m\geq1$, see below. One quarter of
the interface is represented as a graph $h(x)=(x+\Lambda)^m$, where
$-\Lambda+\varepsilon\leq x \leq 0$. The second quarter is obtained by
reflecting this graph with respect to the $x$-axis. Then, by reflecting the two
branches with respect to the $y$-axis, we obtained the closed interface we
worked with. In this manner we could exploit the four-fold symmetry of the
domains and achieve a four-fold reduction in the number of algebraic equations,
approximating the integral equation. For $m<1$, the tip is smooth, and we took
$\varepsilon=0$. For $m>1$ there is a cusp at $x=-\Lambda$ that our numerical
method can not handle. A similar difficulty arises for $m=1$ if the wedge angle
is very small. A positive $\varepsilon$ allows one to employ the method, if the
node spacing in the cutoff region is less than $\varepsilon^m$.

We measured the retreat distance of the tip $L(t)$ and the interface shape at
different rimes for three values of $m$: $m=1/4$, $1/2$ and $5/4$, and also for
three different wedge angles for $m=1$. In the process of relaxation, each of
these domains ultimately becomes a perfect circle. Therefore, to observe the
self-similar asymptotics we performed the measurements at times much shorter
than the characteristic time of relaxation toward a circle. In addition, to
minimize the influence of other tips, we performed the measurements sufficiently
close to a chosen tip (at distances much smaller than the distance between the
chosen tip and the neighboring tip). These two limitations are especially
relevant at $m<1$, where theory predicts self-similarity at sufficiently long
times.  On the contrary, at $m>1$ we performed the measurements at very short
times. Here the main limitation comes from the presence of the cutoff, which
necessitates a sufficiently long ``waiting time" so that the influence of the
cutoff on the solution can be neglected.

For $m=1/4$ we took $\Lambda=2 \times 10^4$ and the minimum node spacing $\delta
S=1/2$. The spacing increased exponentially with the distance from the tip, and
the initial number of nodes was $2 \times 10^3$ (here and in the following --
per quarter of the interface). For $m=1/2$ we chose $\Lambda=10^5$, $\delta S=1$
and the initial number of nodes $10^3$. The set of numerical parameters for
$m=5/4$ was $\Lambda=10^{-3},$ $\delta S=10^{-8}$, $\varepsilon=(2\times
10^{-8})^{4/5}=6.93\times 10^{-7}$, and the initial number of nodes $1200$.

For $m=1$ and the wedge angles $\theta=10^{\circ}$ and $5^{\circ}$ we used
$\Lambda=1$, $\delta S=10^{-5}$, and the initial number of nodes $10^3$.  The
cutoff parameter $\varepsilon$ was chosen to be $\varepsilon =3\delta
S\cot\theta\approx0.017$ for $\theta=10^{\circ}$, and $0.034$ for
$\theta=5^{\circ}$. For $\theta=2^{o}$ we took $\Lambda=3\cdot10^{-3}$, $\delta
S=10^{-6}$, the initial number of nodes $2\cdot 10^{3}$, and the cutoff
parameter $\varepsilon\approx 2.8\cdot 10^{-5}$.

In all cases the time step was chosen to be $10^{-3}$ times the maximum of the
ratio of the interface curvature radius and the interface speed at the same
node. This choice of numerical parameters was dictated by the fact that, at
$0<m<1$, we were interested in a long-time behavior, whereas at $m>1$ we needed
to focus on very earlier times, in order to observe the predicted
self-similarity and scalings.

\begin{figure}[ht]
\includegraphics[width=7.5 cm,clip=]{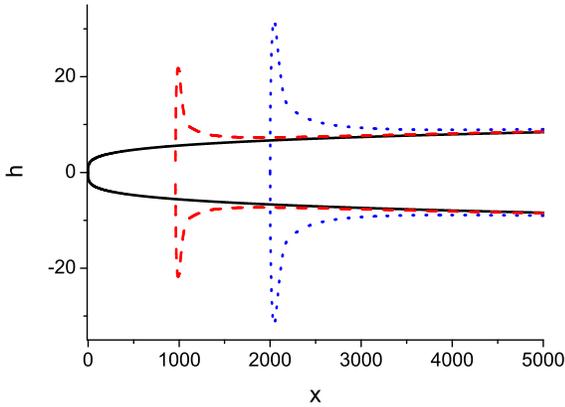}
\caption{Snapshots of a part of the simulated system for $m=1/4$ at times $t=0$
(the black solid line), $2.7 \times 10^{5}$ (the red dashed line) and $1.2
\times 10^{6}$ (the blue dotted line). Notice the large difference between the
horizontal and vertical scales.} \label{snapshots1/4}
\end{figure}

\begin{figure}[ht]
\hspace{0.6cm}
\includegraphics[width=6.0 cm,clip=]{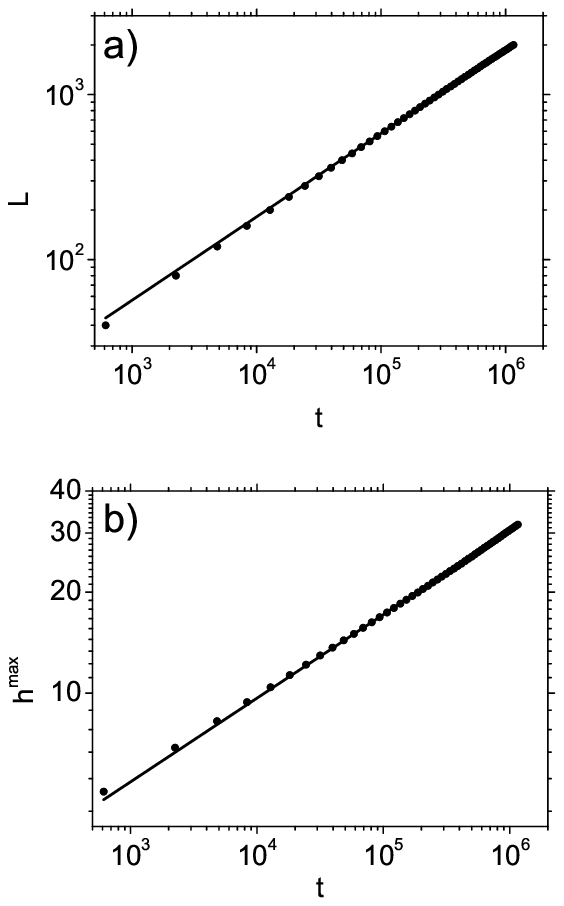}
\caption{Figure a shows, in a log-log scale, the retreat distance $L(t)$ and its
power-law fit $1.73 \,t^{0.50}$ for the case of $m=1/4$. Figure b shows, in a
log-log scale, the local maximum interface elevation, $h^{max}(t)$, and its
power-law fit $0.97\, t^{0.25}$.} \label{lengths1/4}
\end{figure}
\begin{figure}[ht]
\hspace{0.7cm}
\includegraphics[width=6.5 cm,clip=]{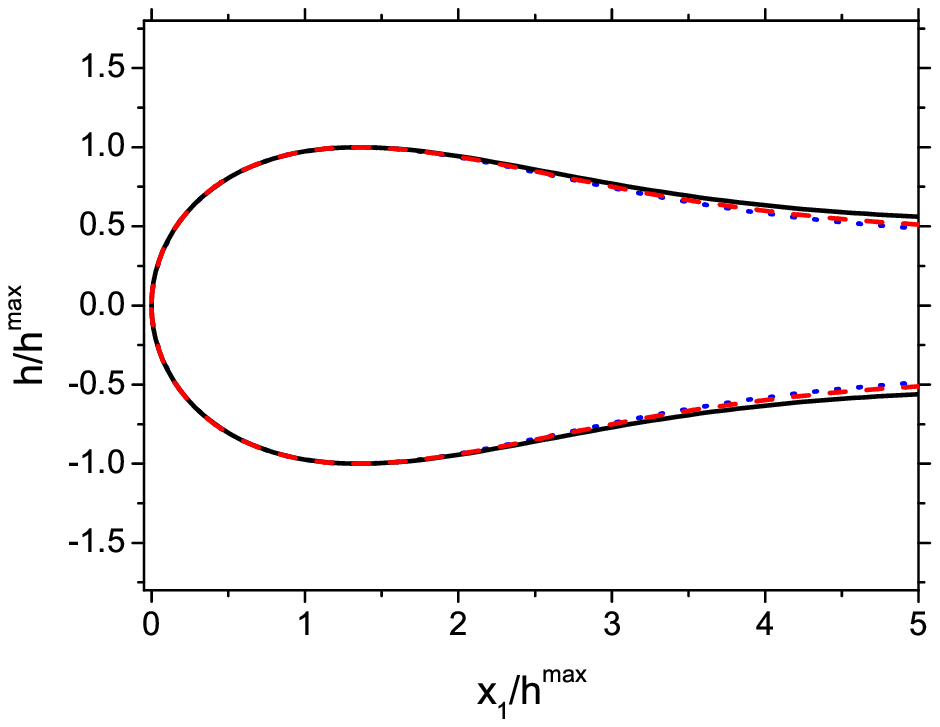}
\caption{Self-similarity at $m=1/4$. Shown is the shape function $h(x_1,t)$,
rescaled to the local maximum elevation $h^{max}$, versus the coordinate $x_1$,
rescaled to $h^{max}$, at times $t=8.35 \times 10^3$ (the black solid line),
$1.07\times10^5$ (the red dashed line) and $1.16 \times 10^6$ (the blue dotted
line).} \label{SS1/4}
\end{figure}

\subsection{Numerical results}
Figure~\ref{snapshots1/4} shows snapshots of a part of the system for $m=1/4$ at
times $t=0$,  $2.7\times 10^5$ and $1.2\times10^6$. One can see a lobe
developing and growing with time. Shown in Fig.~\ref{lengths1/4}a is the retreat
distance $L(t)$ versus time. A power law fit yields exponent $0.50$ which
coincides with the predicted theoretical value $1/2$, see Eq.~(\ref{scalings2}).
It is more convenient numerically to measure the local maximum height of the
interface $h^{max}(t)$, rather than the curvature radius at the tip $R(t)$.
Because of the self-similarity, the quantities $h^{max}(t)$ and $R(t)$ are
expected to exhibit the same power law dependence (of course, with different
pre-factors). Fig.~\ref{lengths1/4}b shows $h^{max}$ versus time in the case of
$m=1/4$. It is seen that this dependence approaches a power law. The fitted
exponent is $0.25$, in excellent agreement with the theoretical value $1/4$.
Figure~\ref{SS1/4} demonstrates the presence of a self-similar region in the
shape function for $m=1/4$. Also noticeable is a rapid (in time) convergence to
the self-similar shape in the lobe region, and a slower convergence in the neck
(the neck can be identified with region II of our theory, as in Ref.
\cite{VMS}).

\begin{figure}[ht]
\hspace{0.4cm}
\includegraphics
[width=7 cm,clip=] {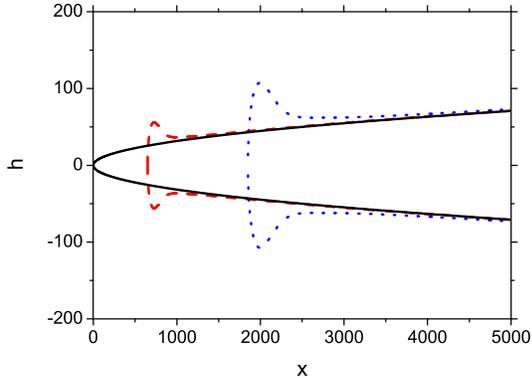} \caption{Snapshots of a part of the simulated
system for $m=1/2$ at times $t=0$ (the black solid line), $1.01 \times 10^6$
(the red dashed line) and $1.1 \times 10^7$ (the blue dotted line). Notice the
large difference between the horizontal and vertical scales.}
\label{snapshots1/2}
\end{figure}

\begin{figure}[ht]
\hspace{0.6cm}
\includegraphics[width=6.0 cm,clip=]{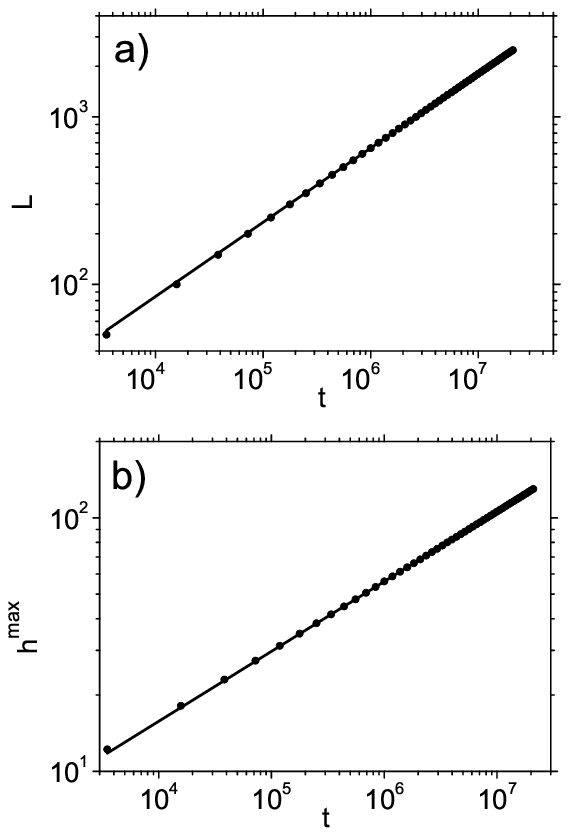}
\caption{Figure a shows, in a log-log scale, the retreat distance $L(t)$ and its
power-law fit $1.43 \,t^{0.44}$ for the case of $m=1/2$. Figure b shows, in a
log-log scale, the local maximum interface elevation, $h^{max}(t)$, and its
power-law fit $1.24\, t^{0.28}$.} \label{lengths1/2}
\end{figure}

\begin{figure}[ht]
\hspace{0.7cm}
\includegraphics[width=6.5 cm,clip=]{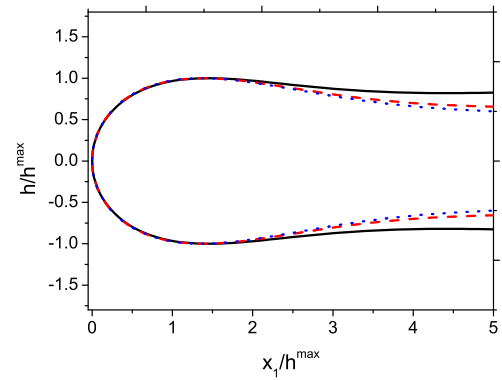}
\caption{Self-similarity at $m=1/2$. Shown is the shape function $h(x_1,t)$,
rescaled to the local maximum elevation $h^{max}$, versus the coordinate $x_1$,
rescaled to $h^{max}$, at times $t=1.57 \times 10^4$ (the black solid line),
$10^6$ (the red dashed line) and $1.06 \times 10^7$ (the blue dotted line).}
\label{SS1/2}
\end{figure}

The numerical results for $m=1/2$ are presented in Figures~\ref{snapshots1/2},
\ref{lengths1/2} and \ref{SS1/2}. Here too power laws for $L(t)$ and $R(t)$ are
observed, and the fitted exponents $0.44$ and $0.28$ are in good agreement with
theoretical values $3/7\simeq 0.43$ and $2/7\simeq 0.29$, respectively. The
shape function again shows self-similarity, with a rapid (in time) convergence
in the lobe region, and a much slower convergence in the neck.

\begin{figure} [ht]
\includegraphics[width=7.5 cm,clip=]{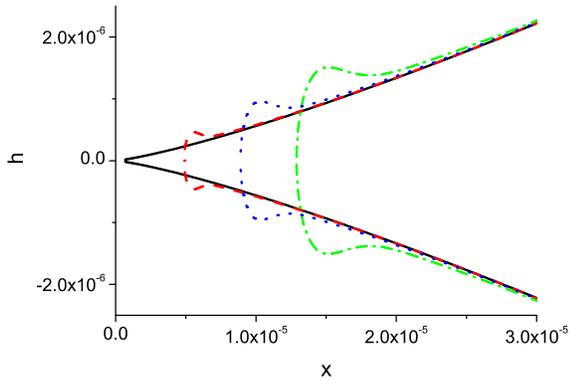}
\caption{Snapshots of a part of the simulated system for $m=5/4$ at times $t=0$
(the black solid line), $3.2 \times 10^{-19}$ (the red dashed line), $2.5 \times
10^{-18}$ (the blue dotted line), and $8.7 \times 10^{-18}$ (the green
dash-dotted line). Notice the large difference between the horizontal and
vertical scales.} \label{snapshots5/4}
\end{figure}

\begin{figure}[ht]
\hspace{0.6cm}
\includegraphics[width=6.0 cm,clip=]{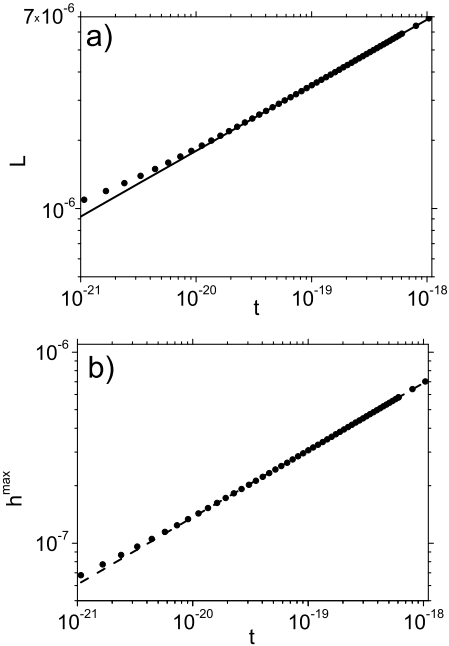}
\caption{Figure a shows, in a log-log scale, the retreat distance $L(t)$ and its
power-law fit $1.1 \,t^{0.29}$ for the case of $m=5/4$. Figure b shows, also in
a log-log-scale, the local maximum elevation $h^{max}(t)$ (the circles) and its
power-law fit $1.6\,t^{0.35}$.} \label{lengths5/4}
\end{figure}

\begin{figure}[ht]
\hspace{0.5cm}
\includegraphics[width=6.5 cm,clip=]{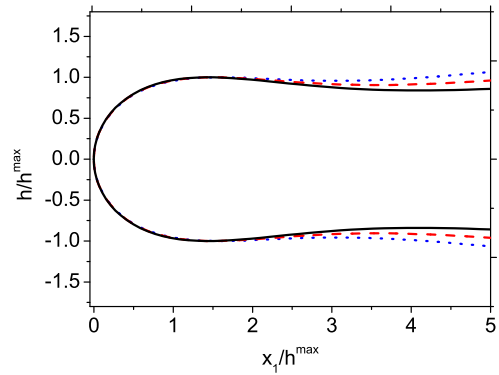}
\caption{Self-similarity at $m=5/4$. Shown is the shape function $h(x_1,t)$,
rescaled to the local maximum elevation $h^{max}$, versus the coordinate $x_1$,
rescaled to $h^{max}$, at times $t=1.1 \times 10^{-19}$ (the black solid line),
$4.9 \times 10^{-18}$ (the red dashed line), and $1.1 \times 10^{-16}$ (the blue
dotted line).} \label{SS5/4}
\end{figure}

The numerical results for $m=5/4$ are shown in Figures~\ref{snapshots5/4},
\ref{lengths5/4} and \ref{SS5/4}. Here the self-similar regime is observed at
very short times. The observed exponents of the power laws ($0.29$ for $\alpha$
and $0.35$ for $\beta$ are in good agreement with theoretical values
$\alpha=0.3$ and $\beta=0.35$. The shape function shows self-similarity in the
lobe region, and a rapid deterioration of self-similarity in the neck as the
time increases.

\begin{figure}[ht]
\hspace{0.5cm}
\includegraphics[width=6.5 cm,clip=]{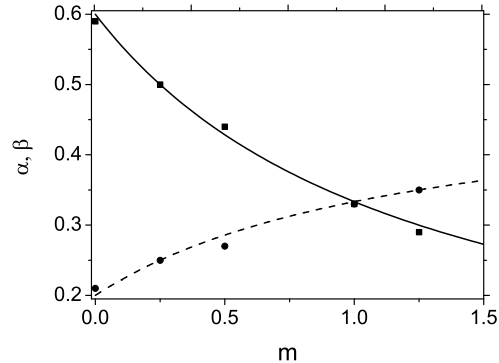}
\caption{The exponents $\alpha$ and $\beta$ of the power laws
(\ref{fullpowerlaws}) versus the parameter $m$. The solid and dashed lines show
theoretical predictions [see Eq. (\ref{scalings2})] for $\alpha$ and $\beta$,
respectively. The corresponding numerical results are shown by the squares and
circles, respectively. The numerical values of $\alpha$ and $\beta$ for $m=1/4$,
$1/2$ and $5/4$ are computed in this work, while those for $m=0$ and $m=1$ are
taken from Refs. \cite{VMS} and \cite{GMV}, respectively.} \label{alphabeta}
\end{figure}

\begin{figure}[ht]
\hspace{0.5cm}
\includegraphics[width=6.5 cm,clip=]{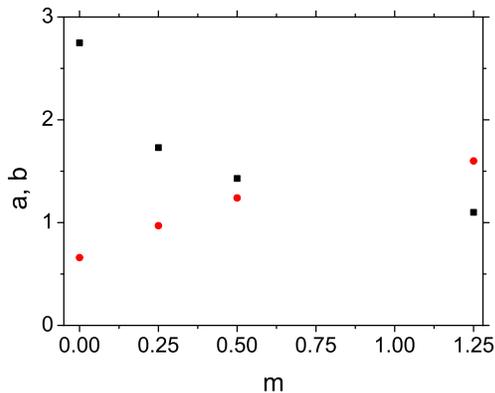}
\caption{The numerically found coefficients $a$ (squares) and $b$ (circles) of
the power laws (\ref{fullpowerlaws}) versus the parameter $m$. The values for
$m=1/4$, $1/2$ and $5/4$ are computed in this work, while those for $m=0$ are
taken from Ref.~\cite{VMS}.} \label{AandB}
\end{figure}

Our numerical results for the dynamic exponents $\alpha$ and $\beta$ at
different $m$ are summarized in Fig.~\ref{alphabeta}. It can be seen that they
follow theoretical curves predicted by Eq. (\ref{scalings2}). The numerically
found coefficients $a$ and $b$ of the power laws are presented in
Fig.~\ref{AandB}.

\begin{figure}[ht]
\hspace{0.5cm}
\includegraphics[width=6.5 cm,clip=]{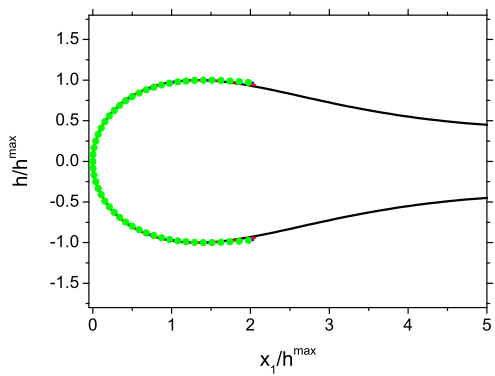}
\caption{Self-similar shape functions at four different values of $m$: $m=0$
(the black solid line), $1/4$ (the red dashed line), $1/2$ (the blue dotted
line) and $5/4$ (the green symbols).} \label{SS1234}
\end{figure}

Figure~\ref{SS1234} depicts, on a single plot, a set of rescaled shape functions
for four different values of $m$. Three of them: for $m=1/4$, $1/2$ and $5/4$,
were computed in the present work, they are the same as shown in
Figures~\ref{SS1/4}, \ref{SS1/2} and \ref{SS5/4}, respectively. The shape
function for $m=0$ is taken from Ref. \cite{VMS}. Remarkably, all the shape
functions coincide in the lobe region. That is, although the retreat distance
and the local maximum elevation of the interface depend on $m$, the rescaled
interface shape is independent of $m$, as long as $m\neq 1$. Figure~\ref{SS1234}
shows it very clearly in the lobe region.  We believe, however, that the
rescaled shape functions actually coincide in the neck too, and that the
self-similar shape function computed for $m=0$ in Ref. \cite{VMS} (see
Fig.~\ref{SS1234}) is valid for \textit{any} $m\neq 1$. Unfortunately, it is
hard to prove this conjecture numerically. We observed that, at $m<1$,
non-self-similar corrections to the self-similar solution in the neck region
decay very slowly with time when $m$ is close to $1$. As a result, one should go
to prohibitively long times (and prohibitively large numerical domain sizes) in
order to reach the ``pure" similarity regime in the neck. In its turn, at $m>1$
corrections to the self-similar solution grow very fast with time when $m$ is
close to 1, so one has to go to prohibitively small times in order to observe
the ``pure" self-similarity in the neck. This is hard to achieve in view of the
presence of the numerical cutoff at $m>1$, which requires sufficiently
\textit{long} waiting times before its influence on the scaling results becomes
small.

\begin{figure}[ht]
\hspace{0.5cm}
\includegraphics[width=6.5 cm,clip=]{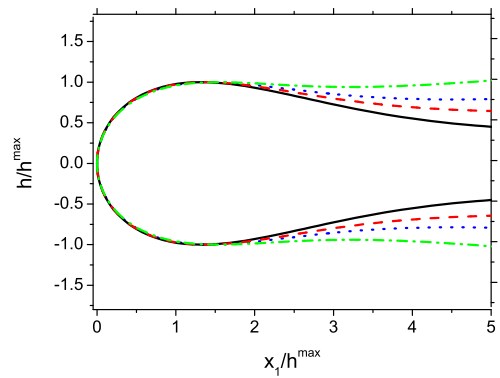}
\caption{Self-similar shape functions for $m=1$ at three different values of the
wedge angle: $10^\circ$ (the green dash-dotted line),  $5^\circ$ (the blue
dotted line), and $2^\circ$ (the red dashed line). The black solid line shows
the shape function for $m=0$.} \label{universal}
\end{figure}

Is there any connection between the universal scaled shape shown in
Fig.~\ref{SS1234} and the rescaled interface shapes of the wedge ($m=1$) which
depend on the wedge angle? To address this question, we simulated the relaxation
dynamics of several wedges with small angles. Figure~\ref{universal} shows
self-similar shape functions for three different values of the wedge angle:
$10^\circ$, $5^\circ$ and $2^\circ$. Shown on the same graph is the shape
function for $m=0$. Remarkably, all the shape functions coincide in the lobe
region. Furthermore, the numerics strongly suggests that, as the wedge angle
tends to zero, the shape function approaches that for $m=0$ \textit{everywhere}.
That is, the observed universal shape function for $m\neq 1$ coincides with the
shape function obtained in the zero-angle limit for $m=1$.

\section{Summary}

We have investigated the dynamics of relaxation, by surface tension, of a family
of curved interfaces, dividing an inviscid and viscous fluids in a Hele-Shaw
cell, and characterizable by a single exponent $m$. A stripe $m=0$, a wedge
$m=1$  and a generic cusp $m=2$ that appears after a pinch-off event,  represent
particular cases of these more general shapes. Combining simple analytic
arguments with a robust numerical method, we have found that, for any $m\neq 1$,
the relaxation dynamics of the interfaces exhibit self-similar intermediate
asymptotics: a late-time asymptotics for $0\leq m<1$ and an early-time
asymptotics for $m>1$. Our theoretical predictions for the dynamic exponents of
the retreat distance and the local maximum elevation of the interface versus
time, at different $m$, are in excellent agreement with numerical simulations.
We have found that, for $m\neq1$, the rescaled interface shape is universal,
that is, it does not depend on $m$ in the similarity region. Remarkably, the
same rescaled interface shape also emerges in the case of $m=1$ (where the
self-similarity is exact and holds for the whole interface at all times $t>0$),
in the limit of zero wedge angle.

Future work should provide a more complete theory that would explain the
surprising findings presented here. We hope that this work (see also Refs.
\cite{VMS,GMV}) will facilitate experimental and further theoretical efforts
aimed at a better understanding of the ``simple" unforced Hele-Shaw flow.

\section*{Acknowledgment}

We thank Omri Gat for a useful discussion. This work was supported by the Israel
Science Foundation (Grant No. 107/05), by the German-Israel Foundation for
Scientific Research and Development (Grant I-795-166.10/2003), and by the
Russian Foundation for Basic Research (Grant No. 05-01-000964).

\pagebreak


\begin{thebibliography}{111}
\bibitem{linear} The damping rates of small sinusoidal
perturbations of flat and circular interfaces are given by the zero-flow-rate
limit of Eq.~(10) of Ref. \cite{ST} (flat interface), and of Eq.~(11) of Ref.
\cite{Paterson} (circular interface).
\end{thebibliography}
\end{document}